\documentclass[12pt]{article}
\usepackage[letterpaper,top=2.6cm,bottom=2.6cm,left=3cm,right=3cm,marginparwidth=1.75cm]{geometry}
\usepackage{amsmath}
\usepackage{amssymb}
\usepackage{amsfonts}
\usepackage{graphicx}
\usepackage{placeins}
\usepackage{lmodern}
\usepackage[font=small]{caption}
\usepackage{booktabs}
\usepackage[numbers]{natbib}
\usepackage{hyperref}
\usepackage{xurl}
\usepackage{longtable}
\usepackage{array}
\usepackage{listings}
\usepackage{csquotes}
\newcolumntype{C}[1]{>{\centering\arraybackslash}m{#1}}
\newcommand{\betamean}{\widehat{\beta_1}}
\newcommand{\sigmean}{\widehat{\sigma_1}}

\title{Correlated Confounding Variables Are Not Easily Controlled for
in Large Survey Research}
\author{William H. Press\\
Computer Science Department and\\
Oden Institute for Computational Engineering and Sciences\\
The University of Texas at Austin}

\begin{document}
\maketitle

\begin{abstract}
    Results in epidemiology and social science often require the removal of confounding effects from measurements of the pairwise correlation of variables in survey data. This is typically accomplished by some variant of linear regression (e.g., ``logistic" or ``Cox proportional"). But, knowing whether all possible confounders have been identified, or are even visible (not latent), is in general impossible. Here, we exhibit two examples that frame the issue. The first example proposes a highly unlikely hypothesis on drug use, draws data from a large, respected survey, and succeeds in ``proving" the implausible hypothesis, despite regressing out more than 20 confounding variables.  The second constructs a ``metamodel" in which a single (by hypothesis unmeasurable) latent variable affects many mutually correlated confounders. From simulations, we derive formulas for the magnitude of spurious association that persists even as increasing numbers of confounders are regressed out. The intent of these examples is for them to serve as cautionary tales.
\end{abstract}

\begin{center}
\small
\subsubsection*{Significance Statement}
\begin{displayquote}
    ``Correlation does not imply causation," is a verity much repeated, even by investigators who disregard it. Still, a correlation between two measurable features can indicate that one {\em may} be in the other's causal chain. However, not all correlations are equally compelling. The association of two variables may be solely due to so-called confounding variables that influence both without the two having any direct relationship. There exist established statistical techniques to control for such confounders. Here, first in a real-data case study and then in an idealized ``metamodel", we show and quantify the pitfalls of such standard techniques. These are cautionary examples of effects understood in theory, but too often ignored in practice.
\end{displayquote}
\end{center}

\section{Introduction}

Since the reader may find the title either tautological or else meaningless across the huge range of possible survey research, some framing and explanation is necessary. Throughout, we use the term ``confounding" in its statistical meaning, namely a variable that influences both the dependent variable and independent variable in a study, causing a spurious association between them that is not evidence of causality.\cite{WikiConfounding}

As an example of the more generally dangerous ``undisclosed flexibility in data collection and analysis" \cite{Simmons}, the pitfalls of compensating (or not compensating) for covariate variables have been widely discussed \cite{Yu}. False positive \cite{Price} or false negative \cite{vanZwieten} results may be produced. General methodologies (e.g., ``deconfounders")  may \cite{WangBlei} or may not \cite{Grimmer} be mitigations. Other methods for inferring causality may be brought to bear \cite{Pearl,Etminan}. This paper joins the cautionary chorus in contributing two very specific examples, one based on actual data, the other on a simple model.

\subsection{Motivating This Paper}

In January, 2025, an Advisory by the U.S. Surgeon General \cite{SurgeonGeneral} flagged alcohol use as ``a leading preventable cause of cancer in the United States." Among statistics put forth in the Advisory was that that the relative risk of breast cancer in women increases by $\sim 10$\% with an additional $\sim 1$ drink per day, and increases further with greater alcohol consumption. These findings were widely reported in the media \cite{Rabin2025,Wapo,reddy2025}. A national network news broadcast summarized one finding as, ``over 16 percent of all breast cancer cases in the U.S. in 2019 were alcohol-related." \cite{PBS2025} The near-simultaneous release of a National Academies report  \cite{NAS2024} on the same subject added some support and some confusion: That study concluded ``with moderate certainty" that alcohol was associated with a higher risk of breast cancer, but that the dose-dependence of the association could be concluded only with ``low certainty".

The Advisory cited studies implicating alcohol in multiple cancer types, including oral cavity, pharynx, larynx, esophagus, breast (women), liver, colon, and rectum. Among the cited studies, one of the largest and most thorough was an Australian study of 226,162 participants that reported hazard ratios and their 95\% confidence limits (roughly, fractional increases in risk) for 26 types of cancer associated with a 7 drink weekly increase in alcohol consumption \cite{Australia}. Of the 26 types, 7 were statistically significant with 95\% confidence and p-values $\le 0.05$ (liver, esophagus, upper digestive tract, pharynx and larynx, colon, breast, colorectum).

\begin{figure}[ht]
\centering
\includegraphics[width=450pt]{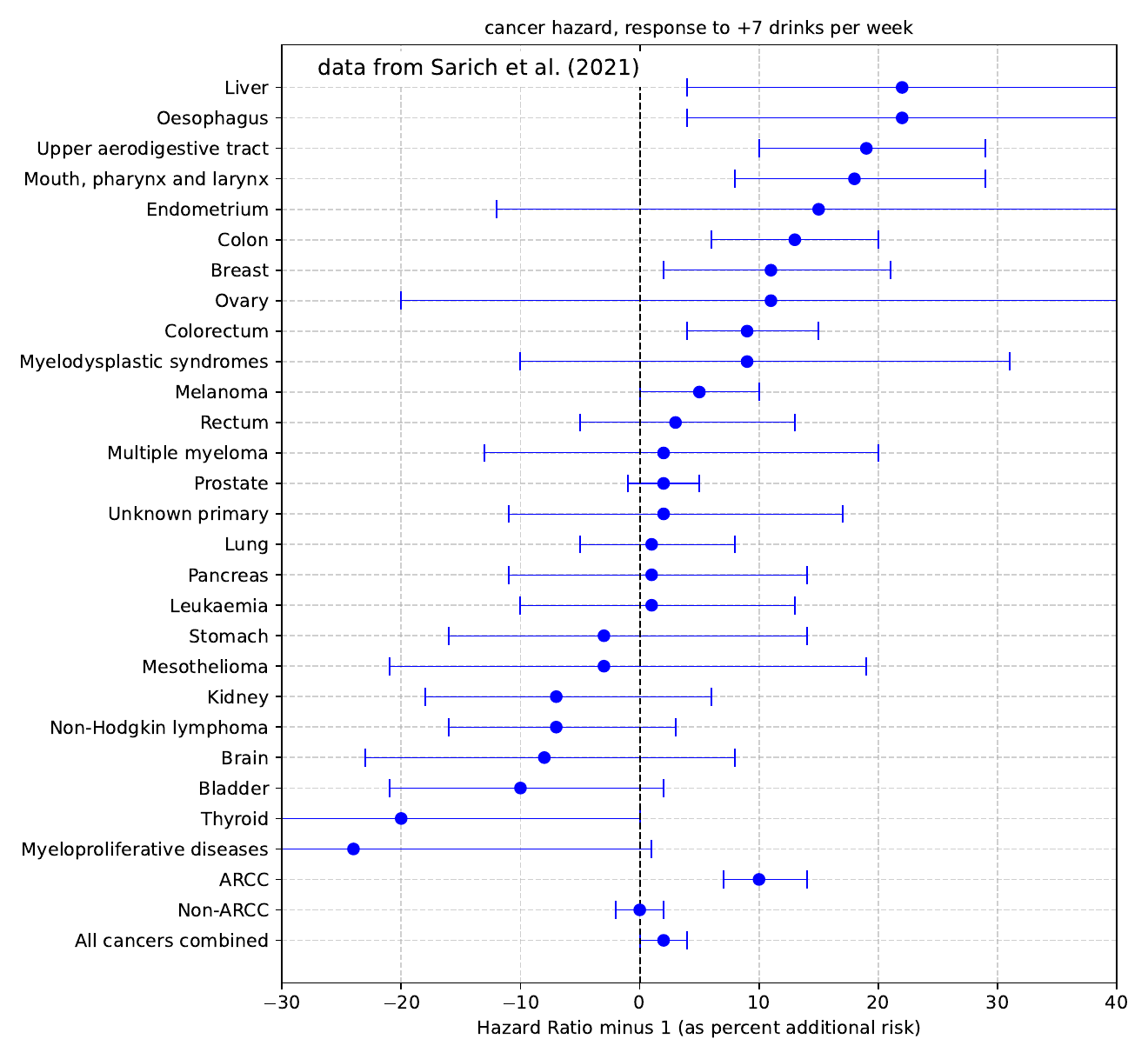}
\caption{From Sarich et al.~(ref.~\cite{Australia}), cancer hazard ratios (roughly, fractional increased risk) associated with increased alcohol use in an Australian study of 226,162 participants. The effects of a large number of potentially confounding variables (see text for partial list) were removed in these authors' analysis by a standard regression technique. The error bars are 95\% confidence intervals.}
\label{fig0}
\end{figure}

Recognizing that various demographic, socioeconomic, and lifestyle
variables might confound measurement of the direct relationship
between alcohol and cancer, the Australia study removed by Cox proportional regression (e.g., \cite{klein}) an extensive list of potentially confounding variables including  sex, education, household income, health insurance status, partner status, country of birth, smoking status and intensity, body mass index, physical activity level, remoteness, and (for individual types of cancers) appropriate subsets of time spent outdoors, skin-tone, diet, parity, menopausal status, hormonal contraception use, and aspirin use. Figure \ref{fig0} shows findings from the Australia study, re-plotted in a format in common with later figures in this paper.

Nothing in this paper is intended to cast doubt on the fact that alcohol consumption causes otherwise preventable cancers. As both the Surgeon General's and National Academies' reports note, causal biochemical pathways are well established: alcohol is metabolized to acetaldehyde, a carcinogen. Alcohol's production of reactive oxygen species can produce inflammation and activate cancer pathways. Alcohol is known to alter hormone levels that are implicated in breast cancer. Other effects are likewise well documented. The Advisory explicitly met the accepted Bradford Hill criteria \cite{WikiBradfordHill} for whether an association merits consideration as causal.

Of concern here is the more general question of whether, in purely survey-based epidemiological and demographic studies, the reporting of derived relative risks for multiple endpoints (here, different cancer types as in Figure \ref{fig0}) should be viewed as quantitatively reliable in the face of removal by regression of many related confounding variables whose causal relationships are likely complex and generally unknown.

Such regressions are not uncommon. A much publicized study of the cancer risk in 92,000 French adults posed by food additives \cite{sellem2024} controlled for age, BMI, height, physical activity, smoking status, number of cigarettes smoked, educational level, family history of cancer, energy intake from lipids and (separately) sugars, sodium, fibre, dairy products; and, for breast cancer, a half dozen further variables. 

There is, of course, no universal answer to the question of reliability. Here, to illuminate some of the issues involved, we construct two examples: The first example proposes a highly unlikely hypothesis on drug use, draws data from a large, respected survey \cite{NSDUH}, and succeeds in “proving” the implausible hypothesis, despite regressing out more than 20 confounding variables. The second constructs a ``metamodel” in which a single (by hypothesis unmeasurable) latent variable affects many mutually correlated confounders.

\subsection{Fictitious Case Study on Real Data}
\label{sec12}

We posit a research project on the misuse of the inhalants listed in Table 1. An investigator entertains the hypothesis that inhalant use is directly attributable to alcohol consumption. Here attributable means not the result of some confounding variable that causes both inhalant and alcohol use, but rather a statistically significant association that remains after all such confounders are accounted for---therefore one that might prove to be causal. While seemingly unlikely, the hypothesis is not immediately dismissible: Perhaps individuals tend to use inhalants preferentially when they are drunk; or perhaps inhalation of alcohol fumes in the course of drinking in some way rewires the brain to favor other inhalant use. At any rate, the investigator sets out to test the hypothesis on actual survey data.

It is an important point that, for this fictitious case study, the hypothesis of direct association is on its face unlikely, while the existence of correlated confounding variables is almost certain. We want to explore circumstances under which accepted methods can produce unreliable or misleading results, in this case statistically significant false positives.

\FloatBarrier

While the investigator's hypothesis may be unlikely, his methodology is assumed to be by-the-book. An ideal data source is the (real, not fictitious) National Survey on Drug Use and Health (NSDUH) \cite{NSDUH} conducted annually by the Substance Abuse and Mental Health Services Administration (SAMHSA), an agency within the U.S. Department of Health and Human Services (HHS). NSDUH 2023 makes available a Public Use File (PUF) with the responses of 56,705 respondents (anonymized as to identity) in 2,636 columns. (Depending on their answers to previous questions, not all respondents are asked all questions.)

\begin{table}[ht]
    \centering
    \small
    \begin{tabular}{lp{2cm}}
        \toprule
        \textbf{Inhalant} & \textbf{Abbreviation} \\
        \midrule
        Amyl nitrite, `poppers,' locker room odorizers, or `rush' & AMYLNIT \\
        Correction fluid, degreaser, or cleaning fluid & CLEFLU \\
        Gasoline or lighter fluid & GAS \\
        Glue, shoe polish, or toluene & GLUE \\
        Halothane, ether, or other anesthetics & ETHER \\
        Lacquer thinner, or other paint solvents & SOLVENT \\
        Lighter gases, such as butane or propane & LGAS \\
        Nitrous oxide or `whippits' & NITOXID \\
        Felt-tip pens, felt-tip markers, or magic markers & FELTMARKR \\
        Spray paints & SPPAINT \\
        Computer keyboard cleaner, also known as air duster & AIRDUSTER \\
        \bottomrule
    \end{tabular}
    \caption{List of Inhalants Studied and Their Abbreviations}
    \label{tab:inhalants}
\end{table}

Among various columns relating to frequency and intensity of alcohol use, the investigator selects as most reliable the response to ``total number of days used alcohol in past 12 months" (column ALCYRTOT), reasoning that self-reported numbers of days is likely to be more accurate than self-reported numbers of drinks. ALCYRTOT, in the range $0\ldots 365$, is thus the independent variable.

The dependent variables of interest are the binary (yes/no) answers to the questions, ``Have you ever, even once, inhaled \_\_\_\_\_ for kicks or to get high?", separately asked for each inhalant listed in Table 1.

Recognizing that confounding variables are likely, the investigator includes in the analysis, and controls for by logistical regression \cite{WikiLogistic,Hosmer}, these potentially confounding responses (see Methods for the NSDUH column names and further details): sex, age, race, marital status, family income, education level, government financial aid status, health insurance status, employment status, household size, urban vs.~rural residence, self-reported overall health, and body mass index. (We refer to this list as ``A".)

\FloatBarrier
\section{Results}

\begin{figure}[ht]
\centering
\includegraphics[width=450pt]{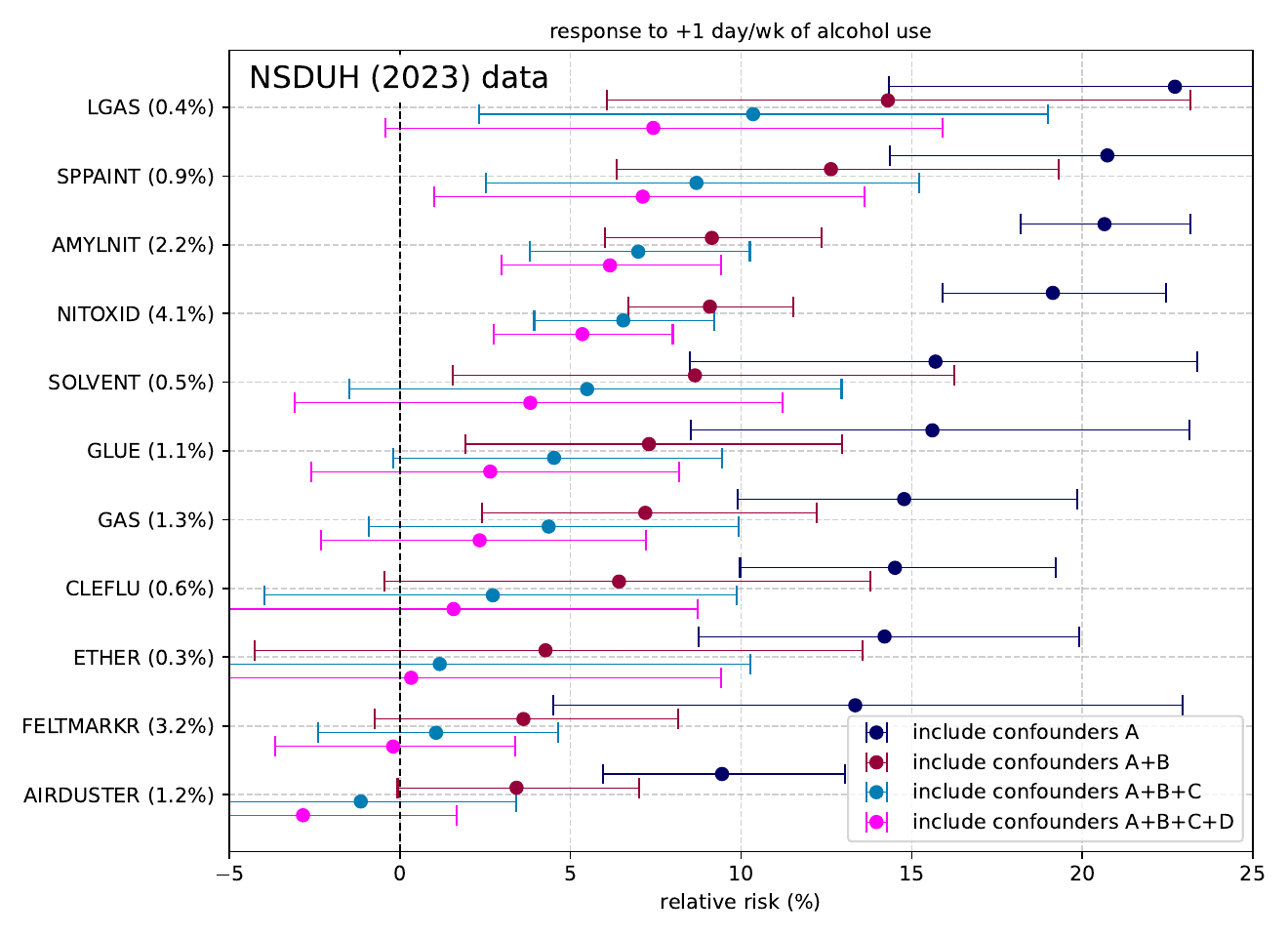}
\caption{Change in the use of inhalants (Table 1) with increase in alcohol consumption, as predicted by logistic regression after correction for multiple potentially confounding demographic and socioeconomic factors (termed ``A", see text). Shown are relative risks associated with one additional day per week of alcohol use. Error bars are 95\% confidence intervals. ``B" adds marijuana use as a confounding variable. ``C" adds 7 hallucinogens as confounders. ``D" adds cigarette use as a confounder. As ``A" through ``D" are successively included in the analysis, the apparent associations with alcohol are successively reduced, most to statistical non-significance. The remaining signals may be a genuine direct association, or may merely indicate that not all confounders are known.}
\label{fig1}
\end{figure}

Figure \ref{fig1} shows the results of the analysis just described, obtained using a standard Python package for logistic regression models (see Methods). Plotted for each inhalant is its so-called relative risk, the fractional change in probability of the dependent variable under a change of (here) a change from zero to one day per week (on average) with alcohol consumption. Percentages listed after the inhalant abbreviations are the baseline reported prevalence of use, so that the relative risk is the fractional increase relative to these values. Error bars shown are 95\% confidence intervals.

Consider first the dark blue values and errorbars, with confounders ``A". One sees a highly statistically significant association between alcohol use and inhalant use for all inhalants tested. Relative risks vary between about 9\% and about 23\%, with strongly significant differences between the low end (felt-tip markers) and the high end (lighter gasses butane and propane). Notable is that these values are found after removal by regression of the long (the investigator argues, exhaustive) list of possible confounders,``A".

We may be more skeptical than the investigator that the included confounding variables are sufficiently exhaustive. Experience may tell us that there are recognizable personality types like ``thrill seeking" or ``addiction prone," not perfectly captured by the socioeconomic and health-related confounders included by the investigator. 

Several columns in the NSDUH questionaire pertain to marijuana use, including (MJEVER) whether ever used, with a 45\% positive response, and ``total number of days used marijuana/cannabis in past 12 months" (MJYRTOT), the exact analog of our independent variable ALCYRTOT for alcohol. We reason that adding these columns to the regression as confounders should correct for addictive (or other related) personality types that could also be associated with alcohol use.

Results are shown as the brown values and errorbars in Figure \ref{fig1}, labeled as ``A+B". We see that indeed all of the relative risks associated with alcohol use are reduced, roughly proportionally, with four now not statistically significant at the 95\% confidence level. Lighter gas, spray paint, amyl nitrite, and nitrous oxide use remain highly significantly associated with alcohol use.

But, we can continue in this manner. Blue values and errorbars labeled ``A+B+C" show the result of adding reported use of each of 7 hallucinogens as confounders (LSD, PCP, peyote, ecstacy/molly, ketamine, dmt/amt/foxy, and Salvia divinorum). The relative risks are all slightly further reduced, again roughly in proportion. Three additional inhalants lose statistical significance. Finally for this analysis, the pink values and errorbars, labeled ``A+B+C+D", show the result of adding cigarette use and intensity in the previous 30 days (CIGEVER and CIG30USE) as confounders, producing additional reductions in relative risk. Lighter gas, originally the largest apparent signal, now loses statistical significance, in part because of large error bars, a consequence of its small prevalence in the population. Howeve spray paints, amyl nitrite, and nitrous oxide remain statistically significant (95\% confidence interval).

\FloatBarrier

We did not identify other columns in NSDUH as plausible confounding variables. A conclusion from this very extensive data set is that spray paints, amyl nitrite, and nitrous oxide inhalent use is directly (i.e., not just via confounding socioeconomic and lifestyle-related variables) associated with alcohol use with high statistical significance, and that several other inhalants may also be, at lower levels of association and with lower statistical confidence.

That may be the conclusion, but should we rely on it? A danger sign is the sequence of reductions in the magnitude of the associations seen as additional confounding variables are introduced into the analysis. Have we overlooked confounding variables that could whittle away the apparent direct association to statistical insignificance? Or, what if such variables exist, but were not included in the survey, hence not accessible to us?

\subsection{Metamodel for Mutually Correlated Confounding Variables}

A metamodel, so called, is a model of models. Each line in Figure \ref{fig1} is a model, specifically the best-fit logistic regression model to a multivariate set of survey variables (including one of primary interest) to a dependent variable. Our metamodel, by contrast, abstracts both the data and the model to a simple set of idealizations. Characteristics seen in the metamodel may then shed light on possible characteristics of real models and their data.

Complex surveys ask questions many of which will show mutual pairwise correlations, not because the one answer directly causes the other, but because both are embedded in some complicated, unknown causal structure. The simplest abstraction of this with (importantly) no actual direct causal relation among the measured variables is, for a population indexed by $i$ with responses $R_{ij}$ indexed by $j$,

\begin{equation}
\begin{aligned}
    \forall i:\quad  &Q_i \in \{-1,1\} \sim \text{Bernoulli }\{0.5,0.5\}\\
    \forall i,j:\quad &R_{ij} \in \{0,1\} = (Q_i Y_{ij} + 1)/2  \; \text{where } Y_{ij} \in \{-1,1\} \sim \text{Bernoulli }\{1-p,p\} 
\end{aligned}
\label{eq1}
\end{equation}
with $p$ a fixed probability in the range $0.5 < p < 1$. Putting this in words, each respondent randomly carries a Bernoulli latent variable;
and that respondent's responses to all questions randomly agree with their latent variable with probability $p > 0.5$, that is, more often than not but still randomly.

The pairwise correlation $r$ between any two responses is readily calculated to be
\begin{equation}
    r=(2p-1)^2 = b^2, \quad b \equiv 2 p -1
\label{eq3b}
\end{equation}
 where where $-1 < b < 1$, parametrizes $p$ as a quantity termed ``bias". Notice that even fairly large values of $p$, say $0.75$, yield only small values of $r$, $0.25$ and even smaller values of $r^2$, the fraction of one variable's variance explained by another, $0.0625$, which would be barely noticeable in a scatter plot of the two variables. While equation \eqref{eq1} posits all positive mutual correlations, one can easily extend the model to have two groups of columns, with mutually positive correlations within each group and negative correlations cross-groups.

Selecting any one of its equivalent columns, label it 0, as a dependent variable of interest (``inhalant use"), and any other column as the independent variable or predictor of interest (``alcohol use"), label it 1, the metamodel finds the best-fitting logistic regression for explaining the former by the latter. Recognizing the existence some number $k-1$ of confounding variables, their columns (2 through $k$) are also included in the fitted model,
\begin{equation}
    \text{logit }(p_i) \equiv \log\left(\frac{p_i}{1-p_i}\right)
    = \beta_0 + \beta_1 R_{i1} + \sum_{j=2}^{k} \beta_j R_{ji}
\label{eq3}
\end{equation}
where $p_i = \text{Prob }(R_{0i}= 1)$ predicts the dependent variable 0 for respondent $i$. The coefficient $\beta_1$ of the predictor in the fitted model, commonly termed the logit coefficient or log-odds coefficient, also with its standard error $\sigma_1$, is the output of one realization of the metamodel.

We mention in passing that logistic regression (equation \ref{eq3}) is different from the Cox proportional regression used in (e.g.) \cite{Australia} and \cite{sellem2024}. However, the two methods are closely related. The Cox model posits a factorable time function, but is otherwise (up to the distinction between log and logit, negligible for small $p_i$'s) a very similar regression for coefficients $\beta_j$,
\begin{equation}
  h(t \mid \{R_{ij}\}) = h_0(t) \exp(\beta_1 R_{i1} + \sum_{j=2}^{k} \beta_j R_{ji}).
\label{eq4}
\end{equation}

In the metamodel, the selected independent variable is, by construction, just another confounding variable. Even in studies of real data, however, regressions akin to equations \eqref{eq3} or \eqref{eq4} allocate their explanatory coefficients $\beta_i$ with independent and confounding variables treated mathematically equivalently; that is the defining idea of ``regressing out" the confounders. Identifying a particular variable as ``independent" is a matter of investigator choice, not mathematics.

We are interested in the average values of $\beta_1$ and its $\sigma_1$ over many realizations of the model, which will depend only the model parameters $p$, $k$, and $N \gg 1$ (the number of respondents). We can define these averages,
\begin{equation}
      \mathbb{E}(\beta_1) \equiv \betamean = \betamean(p,k),
      \quad \mathbb{E}(\sigma_1) \equiv \sigmean = \sigmean(p,k,N)
\end{equation}
While it may be possible to characterize these ``universal" functions $\betamean(p,k)$ and $\sigmean(p,k,N)$ analytically in some limits, it is also easy to approximate them by simulation for any desired values. Doing so (see Methods) we have found purely empirical functional forms that are adequate approximations for present purposes,
\begin{equation}
    \begin{aligned}
        \betamean(p,k) &\approx \frac{3 b^2}{k}\\
        \sigmean(p,k,N) &\approx N^{-1/2} (4 + 12\,b^5)\left(\frac{k-(1+b)/4}{k}\right)        
    \end{aligned}
\label{eq5}
\end{equation}

\begin{figure}[ht]
\centering
\includegraphics[width=450pt]{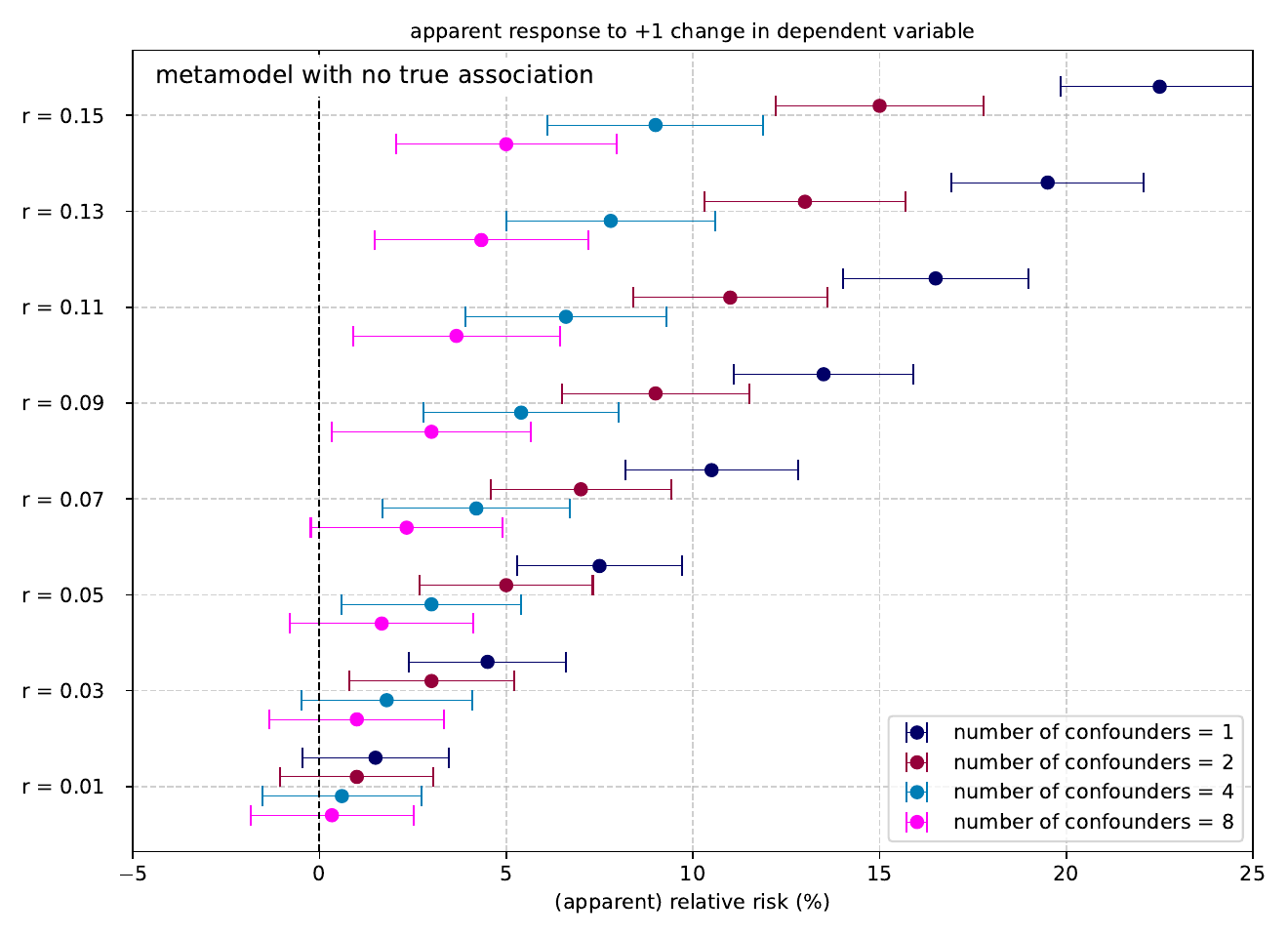}
\caption{Results from metamodel described in the text. Every member of a large fictitious population of size $N$ has multiple binary features, none causally related, but all mutually correlated with constant pairwise correlation coefficient $r$. An investigator takes one feature as the dependent variable and seeks to measure the direct influence of another feature on it, not knowing that the true answer is zero. Recognizing correlated variables as confounders, the investigator removes $n=1, 2, 4,$ or $8$ of them by regression. The figure shows the measured direct association (plotted as relative risk as in Figure \ref{fig1}) for various values of $r$ and $n$. Error bars are 95\% confidence intervals for $N\approx 50,000$, as in Figure \ref{fig1}, but would scale as $N^{-1/2}$.}
\label{fig2}
\end{figure}

Figure \ref{fig2} displays the results of equations \eqref{eq5} for a range of values of mutual correlation $r$ (equivalent to ranges of $p$ or $b$, see equation \ref{eq3b}) and number of confounders ($k-1$) in essentially the same format as Figure 1, and with the same $x$-axis scale. Errorbars in Figure 2 are normalized to the same population size as Figure 1. In the ranges of $r$ and $k$ shown one sees a pattern strikingly similar to that in Figure 1, despite the fact that the metamodel has no actual direct causal associations, only confounding variables with mutual correlations $r$ ranging from 0.01 to 0.15. 

\FloatBarrier

\section{Discussion}

Figures \ref{fig1} and \ref{fig2} are not directly comparable line-by-line: Where the metamodel assumes a single causal latent variable yielding a single value of mutual correlation among many observables, the variables in actual survey data are related by a complicated causal web of many observable and latent variables. Still, the lines in the metamodel may correspond to magnitudes of confounding in some rough average sense.

\begin{figure}[ht]
\centering
\includegraphics[width=450pt]{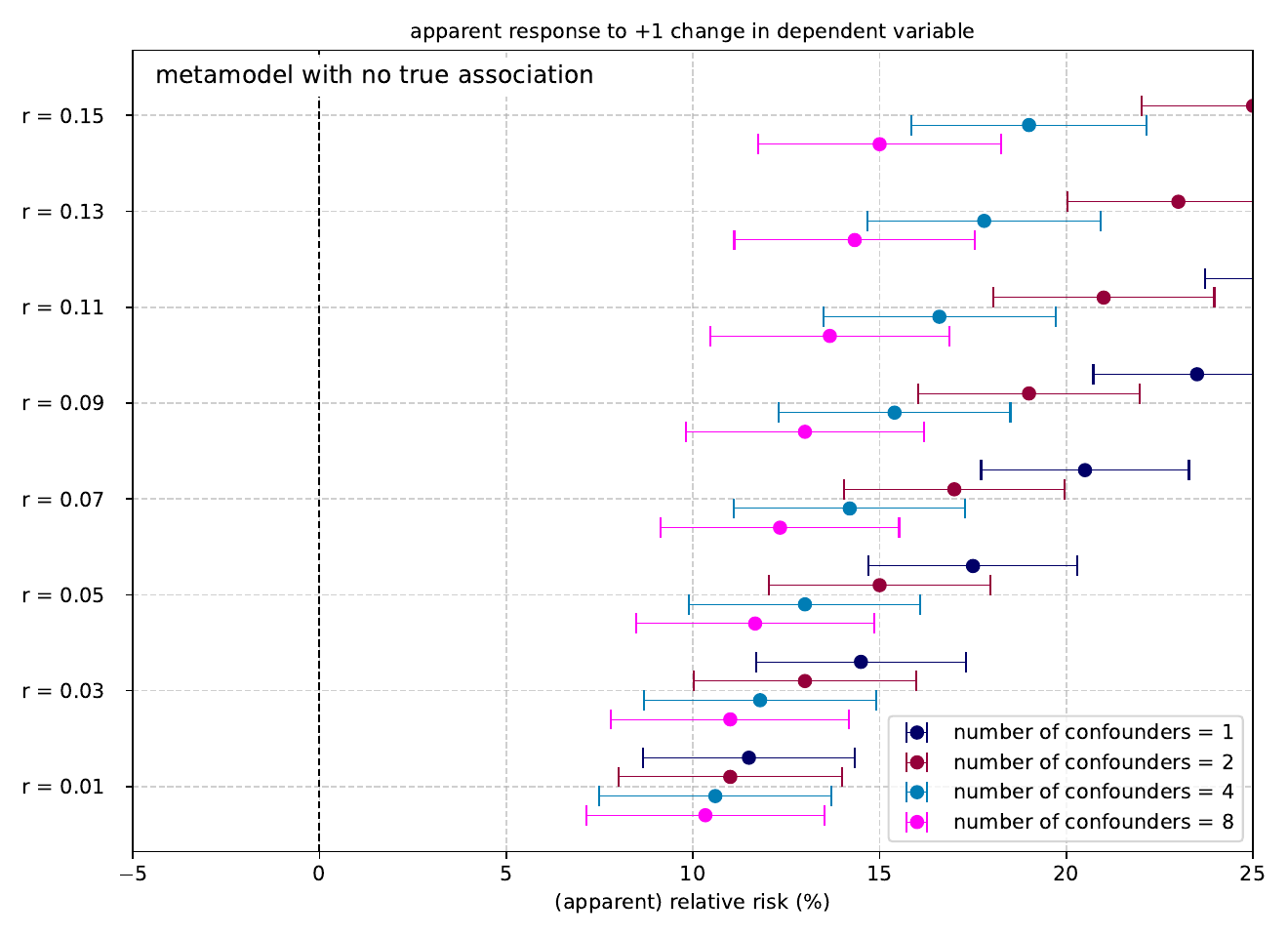}
\caption{Like Figure \ref{fig2}, except that a value 0.10 is added to the logit probability in drawing values of the dependent variable. }
\label{fig3}
\end{figure}

Figures \ref{fig1} and \ref{fig2} are both cases where the measured direct associations are spurious---by construction in the metamodel and with near certainty for the inhalant hypothesis. What should we expect to see in the case of a true direct association? We can extend the metamodel to give some indication of this. Instead of drawing values for the dependent variables $R_{i0}$ with correlated probabilities $p_i$, we can draw them with probabilities
\begin{equation}
    p_i^\prime = \text{logit}^{-1}[\text{logit}(p_i)+\beta^\prime R_{i1}]
\end{equation}
which makes the dependent variable (index 0) causally related to the independent variable (index 1) by a logit increment $\beta^\prime$. Figure \ref{fig3} shows the results for $\beta^\prime = 0.10$. As expected, Figure \ref{fig3} and Figure \ref{fig2} differ in first order by an increase of 10\% (0.10) in all the relative risks, with also some changes in the size of error bars. Worth noting is that confounding variables with larger mutual correlations (top line in Figure \ref{fig3}) significantly bias the observed relative risk to larger than its true (i.e., causal) value.

While the coefficients $\beta_i$ in equations \eqref{eq3} or \ref{eq4}) enter the models in linear combination, there is no reason to think that nonlinear regressions (e.g., \cite{nonlinear}) would produce qualitatively different findings. Any type of regression will allocate some weight to both the desired independent variable and any included confounding variables that have a statistical association (linear or not) with the dependent variable. And, as in the cases considered above, the independent variable may be just another confounder whose association is not causal.

For actual studies such as \cite{Australia} and \cite{sellem2024}, one could contemplate constructing a series of models that add confounding variables sequentially, one by one or in groups. It is not clear, however, that much would be learned. Convergence to a putative true association could be real, or it could simply indicate that an important latent confounding variable is not present in the survey data. Here, we purposely constructed clean examples of pure confounding.

The general conclusion that can be drawn, however unsurprising, is that the quantitative measurement of small associations in the presence of a complex web of confounding associations is a fraught exercise, to be approached with great caution.

\section{Methods and Materials}

\subsection{Data Source and Curation}
The NSDUH 2023 Public Use File (delimited) \cite{NSDUH} and PUF Codebook \cite{codebook} were downloaded from \url{https://www.samhsa.gov/data/data-we-collect/nsduh-national-survey-drug-use-and-health/datafiles}.

Seventy-nine columns were identified as potentially interesting (see SI for list). For each column, responses that were not natively ordinal (i.e, logically ordered) were mapped to the most logical or most neutral ordinal value. For example, 99 for No Response might, for most questions, be mapped to 0, for ``none" or ``no". Categorical variables (e.g., race) were mapped to consecutive integers from 0 and tagged for one-hot encoding. See SI for the complete list of mappings.

\subsection{Model Fitting}

A single column, one of those listed in Table 1, was chosen to be the dependent variable. The independent variables, all treated equivalently, were ALCYRTOT plus, for the four cases shown in Figure \ref{fig1}, the confounding variable columns listed in Table 2.

\begin{table}[h]
    \centering
    \renewcommand{\arraystretch}{1.5} 
    \begin{tabular}{|C{1cm}|p{10cm}|} 
        \hline
        Case & Confounding Variables Included \\ 
        \hline
        A & IRSEX, ANYHLTI2, GOVTPROG, IREDUHIGHST2, IRFAMIN3, BMI2, HEALTH2, AGE3, COUTYP4, IRHHSIZ2 (demographic and health variables)\\ 
        \hline
        B & all A, plus MJYRTOT, MJEVER (marijuana-use variables)\\ 
        \hline
        C & all A and B, plus LSD, PCP, PEYOTE, ECSTMOLLY, KETMINESK, DMTAMTFXY, SALVIADIV (hallucinogen-use variables)\\ 
        \hline
        D & all A, B, and C, plus CIGEVER, CIG30USE (cigarette-use variables)\\ 
        \hline
    \end{tabular}
    \caption{Confounding variables included in the analyses A, B, C, and D shown in Figure \ref{fig1}. (For variable-name decoding, see NSDUH Codebook at \cite{codebook}.)}
    \label{tab:example}
\end{table}

The logistic regression model best fitting the data was in all cases found by the standard package ``statsmodels" \cite{statsmodels}, coded as,
{\tt\begin{verbatim}
    import statsmodels.api as sm
    model = sm.Logit(arr[:,0], arr[:,1:k+1])
    result = model.fit(disp=0)
    coef = result.params
    std_err = result.bse
\end{verbatim}}

That the returned coefficients are themselves relative risks
for a dependent variable upon changing a single independent variable by one unit can be seen by
\begin{equation}
\begin{aligned}
    \text{relative risk} &\equiv \frac{p^\prime}{p} - 1
    = \frac{\text{logit}^{-1}[\text{logit}(p)+\beta_1]}{\text{logit}^{-1}[\text{logit}(p)]} - 1\\
    &= \frac{\exp(\beta_1)}{1+[\exp(\beta_1)-1]\,p} - 1 \approx \beta_1 (1-p)
    \approx \beta_1\; ,
\end{aligned}
\end{equation}
where the approximate equalities are for $\beta_1 \ll 1$ and $p \ll 1$.

\subsection{Empirical Fitting Formulas for the Metamodel}

Five hundred synthetic data sets of size $N=10000$ were generated according to equation \eqref{eq1} and then fit by the code above (same code as used for the real data). The resulting logit coefficients were averaged over equivalent columns and all data sets and plotted as functions of $p$ and $k$. (See SI for these plots.) The simple fitting formulas, equation \eqref{eq5}, were devised by trial and error. SI Figures \ref{figSI2} and \ref{figSI2} show the simulation and fitting formula results.

\FloatBarrier

\subsection*{Acknowledgments}
I am grateful to David Donoho for a close reading of this paper and for suggesting multiple references to the literature. Donoho also outlined a more rigorous, purely analytical approach to several issues treated above only by simulation, an elegant approach that would be his work, not mine, hence not in scope for this paper.

\bibliographystyle{pnas-new}
\bibliography{sample.bib}

\renewcommand\thefigure{SI.\arabic{figure}}    
\section*{Supporting Information}
\setcounter{figure}{0}    

\subsection*{Data Source and Curation}

The columns selected as potentially interesting and their response mappings (see main text) are given in the following table. As an example, the entry ``ALCEVER ORD 2:0, 85-97:0" means that ALCEVER, with tabulated responses 1 (yes), 2 (no), 85 (bad data), 94 (don't know), and 97 (refused) are mapped by us to the ordinal values 0 (no or other) and 1 (yes). For variable-name decoding, see NSDUH Codebook at the main text's ref.~\cite{codebook}.

\begin{longtable}{|l|l|l|}
    \hline
    Column & Type & Mappings \\ 
    \hline
    \endfirsthead
    
    \hline
    Column & Type & Mappings \\ 
    \hline
    \endhead

    \hline
    \multicolumn{3}{r}{\textit{Continued on next page}} \\ 
    \hline
    \endfoot
    
    \hline
    \endlastfoot
    
        IRSEX & CAT & 1:0, 2:1 \\ 
        IREDUHIGHST2 & ORD &  \\ 
        EDUHIGHCAT & ORD & 5:0 \\ 
        IRFAMIN3 & ORD &  \\ 
        ANYHLTI2 & ORD & 2:0, 94-98:0, 1:1 \\ 
        IRMARIT & CAT & 99:5 \\ 
        CIGEVER & ORD & 2:0, 1:1 \\ 
        CIG30USE & ORD & 91-98:0 \\ 
        CIG30AV & ORD & 91-98:0 \\ 
        BMI2 & ORD &  \\ 
        COUTYP4 & ORD & 3:0, 2:1, 3:2 \\ 
        AGE3 & ORD &  \\ 
        NEWRACE2 & CAT & 3-4:3, 5:4, 6:5, 7:6 \\ 
        ALCEVER & ORD & 2:0, 85-97:0 \\ 
        ALCTRY & ORD & 985-998:80 \\ 
        ALCYRTOT & ORD & 985-998:0 \\ 
        ALCDAYS & ORD & 85-98:0 \\ 
        ALCUS30D & ORD & 975:4, 985-998:0 \\ 
        ALCBNG30D & ORD & 80-98:0 \\ 
        IRALCFY & ORD & 991-993:0 \\ 
        HEALTH & ORD & 94-97:3 \\ 
        HEALTH2 & ORD &  \\ 
        IRWRKSTAT18 & CAT & 99:0 \\ 
        IRHHSIZ2 & ORD &  \\ 
        IRKI17\_2 & ORD &  \\ 
        GOVTPROG & ORD & 2:0 \\ 
        NICVAPEVER & ORD & 2:0, 94-97:0 \\ 
        NICVAPAGE & ORD & 985-998:80 \\ 
        NICVAP30N & ORD & 91-98:0 \\ 
        ILLYR & ORD &  \\ 
        SNYATTAK & ORD & 85-99:1 \\ 
        SNRLGSVC & ORD & 85-99:0 \\ 
        SNRLGIMP & ORD & 1-2:0, 85-99:1, 3-4:2 \\ 
        IRDSTNRV30 & ORD & 99:5 \\ 
        IRDSTHOP30 & ORD & 99:5 \\ 
        IRIMPSOC & ORD & 99:0 \\ 
        HTINCHE2 & ORD & 985-998:66 \\ 
        WTPOUND2 & ORD & 9985-9998:150 \\ 
        MJEVER & ORD & 2:0, 94-97:0 \\ 
        MJDAY30A & ORD & 91-98:0 \\ 
        MRJYR & ORD &  \\ 
        IRMJFY & ORD & 991-993:0 \\ 
        BLNTEVER & ORD & 2:0, 4:0, 11:1, 85-98:0 \\ 
        COCEVER & ORD & 2:0, 94-97:0 \\ 
        COCYR & ORD &  \\ 
        CRKEVER & ORD & 2-98:0 \\ 
        CRKYR & ORD &  \\ 
        HEREVER & ORD & 2-97:0 \\ 
        HERYR & ORD &  \\ 
        LSD & ORD & 2-97:0 \\ 
        LSDYR & ORD &  \\ 
        PCP & ORD & 2-97:0 \\ 
        PEYOTE & ORD & 2-97:0 \\ 
        MESC & ORD & 2-97:0 \\ 
        PSILCY & ORD & 2-97:0 \\ 
        ECSTMOLLY & ORD & 2-97:0 \\ 
        KETMINESK & ORD & 2-97:0 \\ 
        DMTAMTFXY & ORD & 2-97:0 \\ 
        SALVIADIV & ORD & 2-97:0 \\ 
        AMYLNIT & ORD & 2-97:0 \\ 
        CLEFLU & ORD & 2-97:0 \\ 
        GAS & ORD & 2-97:0 \\ 
        GLUE & ORD & 2-97:0 \\ 
        ETHER & ORD & 2-97:0 \\ 
        SOLVENT & ORD & 2-97:0 \\ 
        LGAS & ORD & 2-97:0 \\ 
        NITOXID & ORD & 2-97:0 \\ 
        FELTMARKR & ORD & 2-97:0 \\ 
        SPPAINT & ORD & 2-97:0 \\ 
        AIRDUSTER & ORD & 2-97:0 \\ 
        METHAMEVR & ORD & 2-97:0 \\ 
        METHAMYR & ORD &  \\ 
        FENTANYYR & ORD &  \\ 
        OXCNNMYR & ORD & 2-98:0 \\ 
        TRQNMYR & ORD &  \\ 
        STMNMYR & ORD &  \\ 
        SEDNMYR & ORD &  \\ 
        OPINMYR & ORD &  \\ 
        MJYRTOT & ORD & 985-998:0 \\ 
        \hline
\end{longtable}

\subsection*{Metamodel Simulation Results}

\begin{figure}[ht]
\centering
\includegraphics[width=440pt]{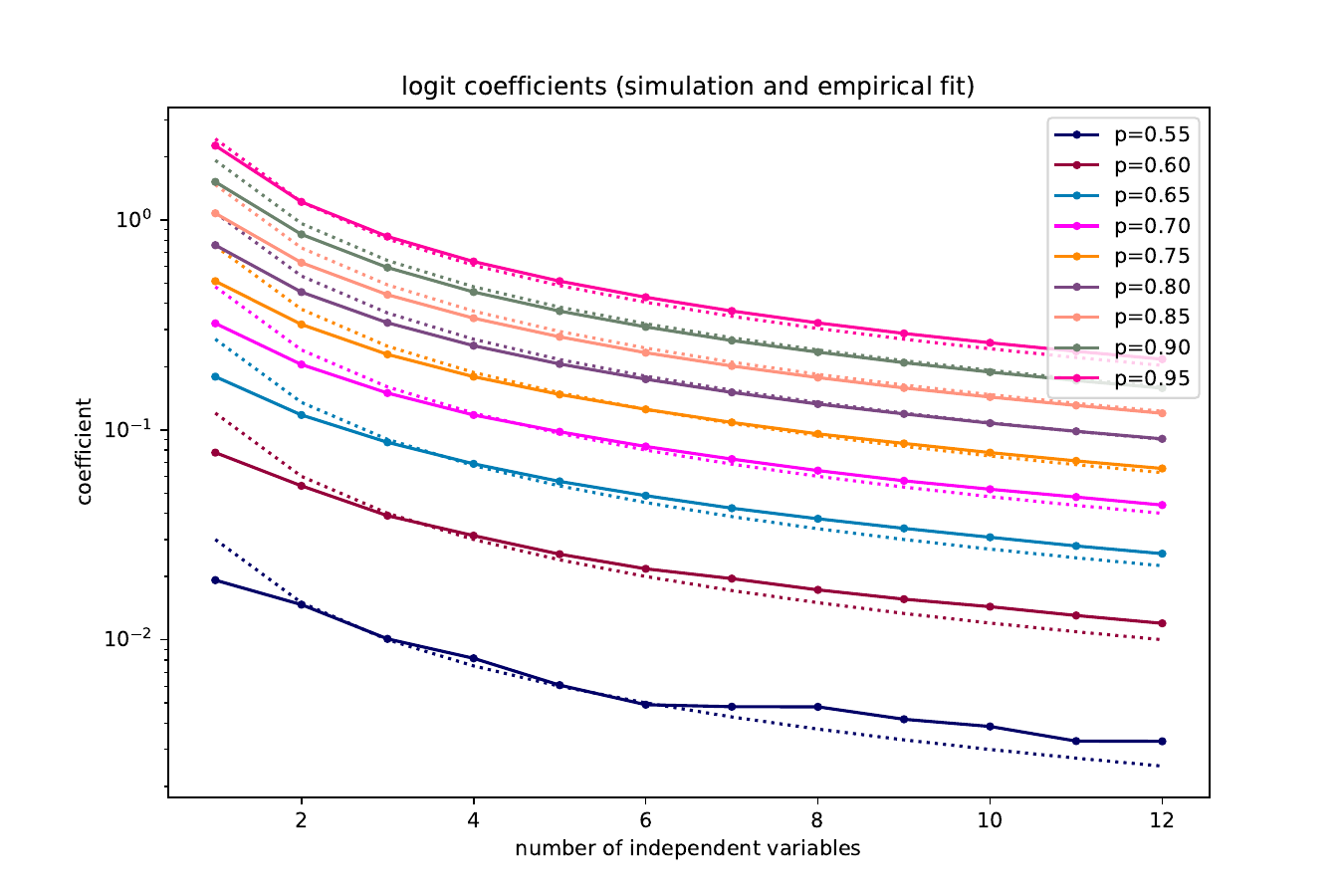}
\caption{Logit coefficient of apparent relative risks in the metamodel and (dotted) empirical fitting formula, main text equation \eqref{eq5}.}
\label{figSI1}
\end{figure}

\begin{figure}[ht]
\centering
\includegraphics[width=440pt]{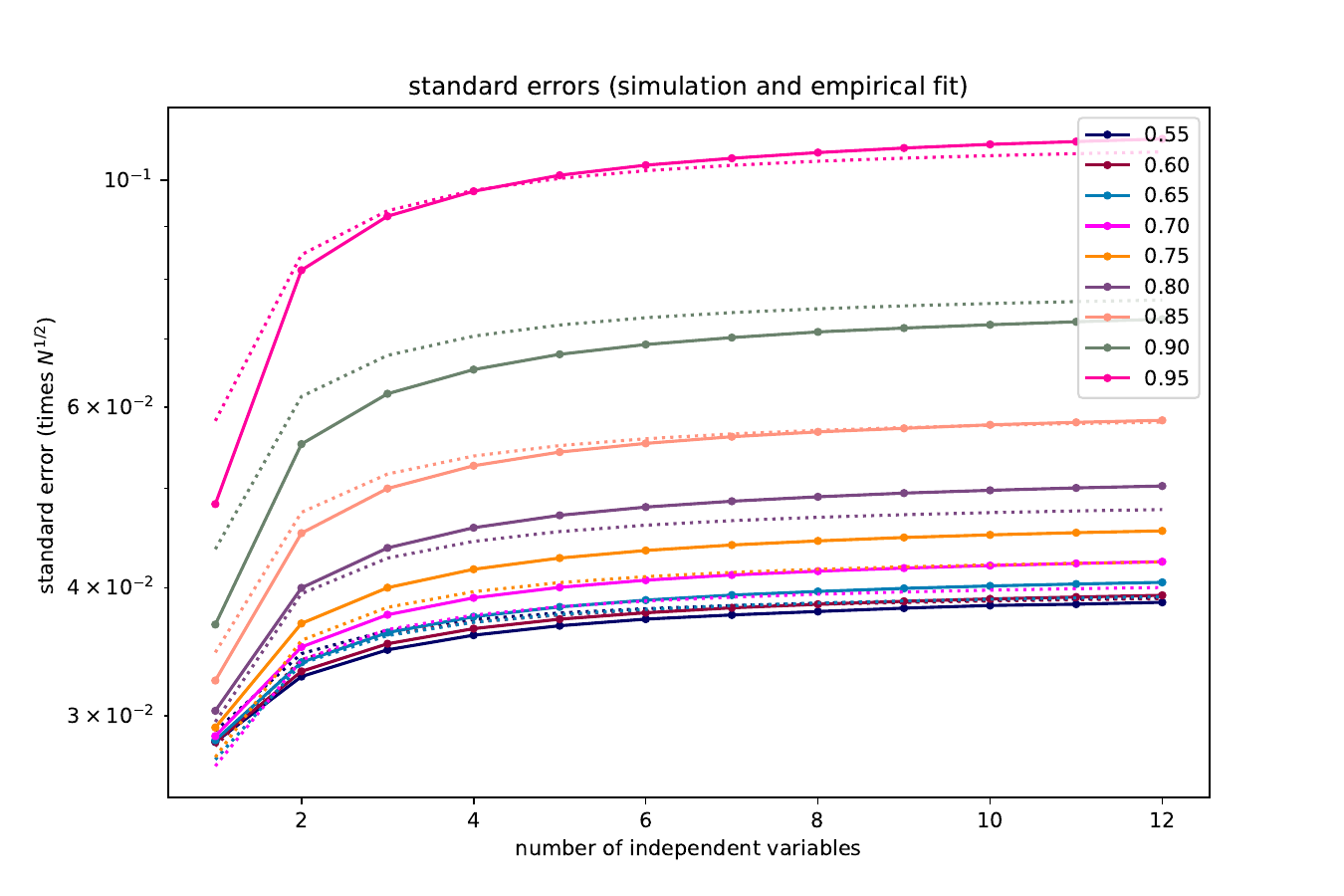}
\caption{Standard error of logit coefficient of apparent relative risks in the metamodel and (dotted) empirical fitting formula, main text equation \eqref{eq5}.}
\label{figSI2}
\end{figure}

\end{document}